\documentclass[sigconf, nonacm]{acmart}

\usepackage[linesnumbered, ruled, vlined]{algorithm2e}
\usepackage{algpseudocode}
\usepackage{makecell}
\usepackage{multirow}
\usepackage{gensymb}
\usepackage{amsmath}
\usepackage{booktabs,caption, threeparttable}
\usepackage[linesnumbered, ruled, vlined]{algorithm2e}
\usepackage{algpseudocode}
\usepackage{diagbox}
\usepackage{booktabs} 
\usepackage{tikz}
\usepackage{pgfplots}
 \usepackage[capitalise]{cleveref}
 \usepackage{subcaption}
\graphicspath{ {Figures/} }
\AtBeginDocument{%
  \providecommand\BibTeX{{%
    \normalfont B\kern-0.5em{\scshape i\kern-0.25em b}\kern-0.8em\TeX}}}
\begin{document}

\title{ReMatch: Retrieval Enhanced Schema Matching with LLMs}

\author{Eitam Sheetrit}
\authornote{Equal contribution.}
\affiliation{%
  \institution{Microsoft}
  \country{Israel}}
\email{eitams@microsoft.com}

\author{Menachem Brief}
\authornotemark[1]
\affiliation{%
  \institution{Microsoft}
  \country{Israel}}
\email{menibrief@microsoft.com}

\author{Moshik Mishaeli}
\authornotemark[1]
\affiliation{%
  \institution{Microsoft}
  \country{Israel}}
\email{mmishaeli@microsoft.com}

\author{Oren Elisha}
\affiliation{%
  \institution{Microsoft}
  \country{Israel}}
\email{oren.elisha@microsoft.com}

\begin{abstract}
Schema matching is a crucial task in data integration, involving the alignment of a source schema with a target schema to establish correspondence between their elements. This task is challenging due to textual and semantic heterogeneity, as well as differences in schema sizes. Although machine-learning-based solutions have been explored in numerous studies, they often suffer from low accuracy, require manual mapping of the schemas for model training, or need access to source schema data which might be unavailable due to privacy concerns. In this paper we present a novel method, named ReMatch, for matching schemas using retrieval-enhanced Large Language Models (LLMs). Our method avoids the need for predefined mapping, any model training, or access to data in the source database. 
Our experimental results on large real-world schemas demonstrate that ReMatch is an effective matcher. By eliminating the requirement for training data, ReMatch becomes a viable solution for real-world scenarios. 

\end{abstract}


%



\maketitle

\section{Introduction}
\label{sec:introduction}
\begin{figure*}
\begin{center}
\includegraphics[width=2\columnwidth]{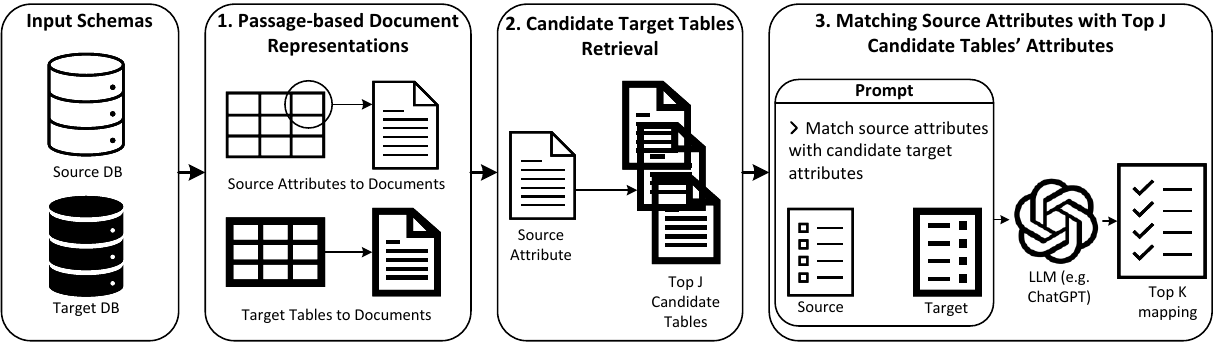}
\end{center}
   \caption{Overview of the ReMatch method.}
\label{fig:fig1-overview}
\end{figure*}
Schema matching is a fundamental task in data management and integration, involving the identification of semantic correspondences between elements of two or more database schemas~\cite{Madhavan2001GenericSM}, regardless of differences in naming, structure, or data type. The matching task is challenging since schemas are designed using different perspectives and terminologies. This semantic heterogeneity can lead to ambiguous mappings where schema elements have the same name but different meanings, or different names but the same meaning.

Human schema matching, a manual and time-consuming process, requires significant effort from skilled individuals. This can be expensive and impractical, particularly in large-scale projects. Furthermore, human matchers are prone to errors and inconsistencies due to cognitive biases and fatigue~\cite{Shraga2021Experts}. As a result, the automation of schema matching has become a major focus within the AI and database-oriented research community over the years.

In recent years, LLMs have achieved significant advancements across many challenging tasks that require a deep understanding of semantics. These models have shown an impressive ability to generalize to new tasks without any task-specific fine-tuning, even in areas significantly divergent from the ones they were originally trained on, including various data related tasks~\cite{Narayan2022CanFM, Zhang2023LargeLM}.

In this paper, we present a new approach that unlike previously proposed machine learning (ML) methods, circumvents the need for predefined mapping, model training, or access to data in the source database. Our approach, which we call ReMatch, improves the task of schema matching by using retrieval-enhanced LLMs.~\cref{fig:fig1-overview} illustrates our proposed approach. 

Our contributions in this study are as follows: (1) Introducing a new method for the task of schema matching, which allows for scalable and accurate matching results, without any model training or access to labeled data. (2) Proposing a mechanism to reduce the search space of the target schema for efficient candidate generation. (3) Exploiting the generative abilities of LLMs to perform semantic ranking between two schemas, in alignment with human matchers. (4) Providing a complete mapping between two important healthcare schemas, aiming to address the lack of real-world evaluation datasets in the field. To the best of our knowledge, this is currently the largest publicly available schema matching dataset.

\section{Background and Related Work}
\label{sec:background}

\textbf{Large Language Models (LLMs)} \quad
Generative language models are trained to generate human-like text and can be fine-tuned for a variety of tasks, but also can be reused for a variety of tasks, with no additional training~\cite{Radford2018ImprovingLU, Brown2020LanguageMA, Liu2023SummaryOC, Mirchandani2023LargeLM}. Embedding models like BERT or Ada~\cite{Neelakantan2022TextAC} provide contextual embeddings that have significantly improved performance on a wide range of NLP tasks~\cite{Sun2019HowTF, Minaee2021DeepLB}. Moreover, the representation of text via embeddings allows for efficient and accurate passage retrieval, using semantic similarity~\cite{Neelakantan2022TextAC, lewis2020retrieval}.

LLMs have been applied to various data preprocessing tasks, including error detection, data imputation, entity matching, as well as schema matching~\cite{Mirchandani2023LargeLM, Narayan2022CanFM}. These models have shown promising results, however, these approaches still faced challenges related to computational expense and inefficiency, as well as lacking evaluation on data that represents real-world schema matching challenges.

\textbf{Schema Matching} \quad
Schema matching involves identifying correspondences between different data schemas. Traditionally, schema matching was done manually, however, this approach strongly relies on the expertise of human matchers~\cite{Dragisic2016UserVI}. This has led to development of automated schema matching, primarily using ML algorithms~\cite{Do2002COMAA, Bernstein2011GenericSM, Madhavan2001GenericSM}. Much of the work on ML-based schema matching was limited to basic schemas, with most works dealing only with toy datasets~\cite{Do2002COMAA, Gal2021LearningTR, Mudgal2018DeepLF, Shraga2020ADnEVCS, Li2020DeepEM}, or focused primarily on entity matching only. This stemmed from a combination of lack of high-quality datasets~\cite{Koutras2020ValentineEM}, as well as relatively poor results of past methods.

Recent work, such as SMAT~\cite{Zhang2021SMATAA} and LSM~\cite{Zhang2023SchemaMU}, leverages advancements in natural language processing to achieve semantic mappings between source and target schemas. While these methods improve previous results, they still require extensive data tagging, limiting their practicality in real-world applications.

Finally, the majority of previous methods dealt with schema-matching as a binary classification task over $\mathcal{A}_1 \times \mathcal{A}_2$. While simplifying training and evaluation, in normalized schemas positive labels scale by $O(n)$, whereas negative labels scale by $O(n^2)$, which is problematic when inference is expensive, as in the case of LLMs.

\section{R\MakeLowercase{e}Match}
\label{sec:matchGPT}

\textbf{Problem Statement} \quad
Given a source schema $\mathcal{S}_1$ and a target schema $\mathcal{S}_2$, with sets of tables $\mathcal{T}_1$ and $\mathcal{T}_2$ and sets of attributes $\mathcal{A}_1$ and $\mathcal{A}_2$, respectively, the schema matching task involves finding a mapping between $(\mathcal{A}_1, \mathcal{T}_1) \in \mathcal{S}_1$ to $(\mathcal{A}_2, \mathcal{T}_2) \in \mathcal{S}_2$.
The matches between their attributes are captured by a relation \textit{match}: $\mathcal{P}(\mathcal{A}_1)  \times \mathcal{P}(\mathcal{A}_2)$. An element $(A_1, A_2) \in$ \textit{match} defines a matching pair possibly representing the same information in the schemas. If $\lvert A_1 \rvert = 1$ and $\lvert A_2 \rvert = 1$, we call the match an \textit{elementary match} or 1:1 match. Otherwise, we refer to it as a \textit{complex match} or m:n match.



Formally, we want to find a function $\Psi: \mathcal{A}_1 \rightarrow \mathcal{P}(\mathcal{A}_2)$, such that $\forall a \in \mathcal{A}_1,\: \forall a'\in \Psi(a),\: (a, a')\in match$. In other words, $\Psi(a) = A'\subseteq \mathcal{A}_2$, should contain all relevant matches for $a$. 

A possible simplification of this problem can be made by limiting $\Psi$ to be a $N \times K, \: N=|\mathcal{A}_1|, K\in \mathbb{N}$, matrix of elements from $\mathcal{A}_2$, denoted as $\Psi_K$. In the case of elementary matches, or m:1 complex matches, the goal from above now becomes to maximize the $accuracy@K$ metric, i.e., to maximize:
\begin{equation}
\label{eq: accuracy_at_k}
    \frac{1}{N} \sum^{N} \mathbf{1}_{\{\exists a', \: (a, a')\in match, \: a'\in\Psi_K(a)\}}
\end{equation}

\textbf{Method Description} \quad ~\cref{algorithm:rematch} shows the pseudocode for our method.
Given a source schema $\mathcal{S}_1$ with a set of tables $\mathcal{T}_1$ and a set of attributes $\mathcal{A}_1$, and a target schema $\mathcal{S}_2$ with a set of tables $\mathcal{T}_2$ and a set of attributes $\mathcal{A}_2$:

\textbf{(1)} Target schema \textit{tables} and source schema \textit{attributes} are transformed into two corpora of structured documents, $\mathcal{C}_t$ and $\mathcal{C}_s$, respectively. Each target schema table and source schema attribute is represented as a structured document consisting of descriptive paragraphs. The title of the document is the table's name, and the opening paragraph provides an overview of the table's purpose and characteristics. The subsequent paragraphs detail the set of attributes serving as the table's primary key, the set of attributes referring to other tables (foreign keys), and the rest of the attributes belonging to this table, respectively. Each attribute is followed by its data type and a textual description. In the case of attribute documents, the specific attribute is highlighted above the title.
    
\textbf{(2)} For each attribute in the source schema, we search for the top $J$ documents that represent candidate tables from the target schema. This step ensures that only the most relevant documents are considered for matching. To facilitate the retrieval of candidate tables, we utilize a text embedding model to encode each source attribute's document, and the corpus of target table documents. These embeddings serve as a basis for measuring semantic similarity, enabling the retrieval of candidate tables. Finally, for every source table $t_i \in \mathcal{T}_1$, we create a set of all top $J$ candidate tables $\mathcal{T}_c$ retrieved. 

\textbf{(3)} In the last step, the LLM is tasked with selecting the top $K$ most similar target attributes from the set of retrieved tables $\mathcal{T}_c$ identified earlier. The model assesses the context provided by the document representations, and yields a ranked list of $K$ potential matches for each attribute in the source schema.

\begin{algorithm}
\caption{ReMatch}\label{algorithm:rematch}
\SetKwInOut{Input}{Inputs}
\SetKwInOut{Output}{Outputs}
\newcommand\mycommfont[1]{\footnotesize\textcolor{blue}{#1}}
\SetCommentSty{mycommfont}
\SetNlSty{textbf}{}{:}
\SetAlgoLined
\DontPrintSemicolon
\Input{ a source schema $\mathcal{S}_1$, $\mathcal{T}_1$, $\mathcal{A}_1$, with all its textual descriptions; a target schema $\mathcal{S}_2$,$\mathcal{T}_2$, $\mathcal{A}_2$ with all its textual descriptions; $K$,$J \in \mathbb{N}$; an embedding model $\Phi$; a generative LLM $\mathcal{F}$.}
\Output{a $N \times K$ matrix of attribute candidates from $\mathcal{A}_2$, denoted as $\Psi_K$.}
$\mathcal{C}_s \gets \emptyset$, $\mathcal{C}_t \gets \emptyset$, $\Psi_K \leftarrow \{\}$\;
\tcc{lines 2-4 are performed once, as a pre-processing step.}
$\mathcal{C}_t \gets TableToDoc(\mathcal{S}_2, \mathcal{T}_2, \mathcal{A}_2)$\;
$\mathcal{C}_s \gets AttributeToDoc(\mathcal{S}_1, \mathcal{T}_1, \mathcal{A}_1)$\;
$\tilde{\mathcal{C}_t} \gets \Phi(\mathcal{C}_t)$\;
\ForEach{$t_i \in \mathcal{T}_1$} {
    $\mathcal{T}_c \leftarrow \{\}$\; 
    \ForEach{$a_k \in \mathcal{A}_1[t_i]$} {
        $\tilde{a}_{emb} \gets \Phi(\mathcal{C}_s[t_i; a_k])$\;
        $scores \gets Similarity(\tilde{a}_{emb}, \tilde{\mathcal{C}_t})$   \;
        $top_j \gets \mathrm{argmaxes}(\{scores_1, \ldots, scores_n\})[1:j]$ \;
        $ \mathcal{T}_c \gets \mathcal{T}_c \cup \mathcal{C}_t[top_j]$
    }
    $\Psi_K \gets CreateTopkMapping(t_i, \mathcal{C}_s[t_i], \mathcal{T}_c, \mathcal{C}_t[\mathcal{T}_c], \mathcal{F})$
}       
\end{algorithm}

\section{Evaluation}
\label{sec:evaluation}
\subsection{Dataset Creation}
\label{subsec:dataset creation}
To evaluate our method we used two primary datasets, both involving mappings between healthcare database schemas. 

\textbf{MIMIC-III to OMOP} \quad
For the first dataset, we created a mapping between the schema of MIMIC-III~\cite{johnson2016mimic}, a public database containing deidentified records from the Beth Israel Deaconess Medical Center, and The Observational Medical Outcomes Partnership Common Data Model (OMOP)\footnote{\url{https://www.ohdsi.org/data-standardization/}}, an open-source data standard. 
The mapping was created manually by a domain expert, aided by the mapping created in~\cite{Paris2021TransformationAE}. If no matching attribute in OMOP could be found, the attribute was assigned \textit{NA}. For convenience we will refer to this dataset as MIMIC.

\textbf{OMAP Benchmark Synthea Dataset} \quad
Following~\cite{Narayan2022CanFM} we used the \textit{Synthea}~\cite{walonoski2018synthea} \textit{to OMOP} dataset from the OMAP benchmark~\cite{Zhang2021SMATAA}. The dataset contains a partial mapping of the schema for Synthea (the source), a synthetic healthcare dataset, to a partial subset of relevant OMOP attributes (the target). For convenience we will refer to this dataset as Synthea.

It is important to note that although both datasets are similar, the first maps \textit{the entire} MIMIC-III schema to OMOP, while Synthea contains only \textit{partial} mappings. To the best of our knowledge, the MIMIC dataset is the largest single-schema to single-schema dataset published. \cref{table: dataset stats} contains statistics detailing both datasets. The full MIMIC dataset is publicly available at~\url{https://github.com/meniData1/MIMIC_2_OMOP}.

\subsection{Experiments}
\label{subsec: experiments}

We established the performance of ReMatch and SMAT~\cite{Zhang2021SMATAA}, the previous state-of-the-art (SOTA) method, on MIMIC and Synthea. 

As our method treats schema matching as a retrieval problem, the most appropriate metric is accuracy@$K$, as defined in~\cref{eq: accuracy_at_k}. Unlike the F1-score, standard accuracy@$K$ is not well defined for m:n matches. To deal with this, we limited ourselves to evaluating 1:1 or m:1 relations. \footnote{It is worth noting that for a dataset with only 1:1 and m:1 mappings, accuracy@1 is equivalent to F1-score when using \textit{argmax}, instead of a threshold, for prediction (since it trivially reflects both the precision and the recall simultaneously).}   

There are multiple dimensions we would like to optimize. The first is maximizing accuracy@$K$. We also seek lower values of $K$, as our aim is to assist human schema matching, and minimize cognitive overload~\cite{Bernstein2011GenericSM, Chernev2015ChoiceOA}. Finally, we would like to minimize the number of target tables, reducing prompt size and costs.

\textbf{Evaluating ReMatch} \quad
For all experiments we used GPT-4~\cite{OpenAI2023GPT4TR}, specifically GPT-4-1106, and Ada2~\cite{Greene_Sanders_Weng_Neelakantan_2022, Neelakantan2022TextAC} as our embedding model. 

We performed a grid search over various values of $J$ and $K$, also allowing us to select hyperparameters for further evaluations. As an ablation study, we evaluated the performance on MIMIC when only the \textit{names} of the tables and attributes are used for retrieval and generation. We also evaluated a version with no retrieval stage, i.e., for each table in the source schema we used the entire target schema. 
Finally, we ran ReMatch multiple times on each of the setups in the ablation study, to verify stability across generations.

\textbf{Evaluating ReMatch with Guidance} \quad
A human matcher may determine that the suggestions are not good enough, or decide to provide information they already have. In our evaluation we investigated a mechanism, inspired by the relevance feedback technique~\cite{Rocchio1971RelevanceFI}. To do so, we selected a mapping of the form $(T_1, a_1) \rightarrow (T_2, a_2)$ for a single attribute from each table. Specifically, we used the \textit{SUBJECT\_ID } attribute where available (19/25 tables), and an attribute containing unique identifiers otherwise. We then automatically included $T_2$ in the set of retrieved tables, $\mathcal{T}_c$, and provided the LLM with this mapping in the prompt.

\textbf{Evaluating SMAT} \quad
Our method's performance was evaluated against  SMAT~\cite{Zhang2021SMATAA}, a previously SOTA non-LLM model. For SMAT's training we used the default hyperparameters: \texttt{\{Learning Rate: 0.8, Batch Size: 64, Epochs: 30\}}, with early stopping, and 10\% of the training data was used as a validation set to avoid over-fitting. Models were trained on a single Nvidia A-100 GPU.

We tried two setups, 80\% training and 20\% testing, and vice versa. The splits were stratified across table names. We also extended the evaluation to include accuracy@$K$. We calculated the accuracy@$K$ by determining whether the correct match received one of the $K$ highest classification scores. Binary predictions were given by using the optimal threshold for the validation set.

\begin{table}
  \caption{Dataset statistics for MIMIC and Synthea. \#Columns and \#Tables refer to the number of columns and tables used in the dataset. \#Mapped Columns refers to the number of unique columns mapped. \#Null Mappings refers to source columns without a mapping (mapped to \textit{NA}).}
  \centering
    \resizebox{\columnwidth}{!}{
  \begin{tabular}{c||c|c|c|c}
    \toprule
\textbf{Dataset}      & \#Columns   &  \#Tables & \#Mapped Columns & \#Null Mappings \\ \hline
MIMIC          &  268 &  25 & 156 & 112\\ \hline
OMOP (MIMIC)   &  425 & 38  & 95  &  - \\  \toprule \bottomrule
Synthea        &  38  &  8  & 105 &  - \\   \hline
OMOP (Synthea) &  67  & 8   & 67  &  - \\ 

\bottomrule
  \end{tabular}}
 \label{table: dataset stats}
\end{table}

\begin{table}
  \caption{Grid search results on MIMIC.}
  \centering
	\resizebox{\columnwidth}{!}{
  \begin{tabular}{c||c|c|c|c|c||c}
    \toprule
Retrieved Documents & Acc@1  & Acc@2  & Acc@3  & Acc@5  & Acc@7  & \textbf{Avg \#T} \\ \toprule
$J=1$                   & 0.424 & 0.589 & 0.644 & 0.709 & 0.697 & 2.44    \\ \hline
$J=2$                   & 0.425 & 0.541  & 0.638  & 0.729 & 0.758 & 4.68    \\ \hline
$J=3$                   & 0.321  & 0.477 & 0.533 & 0.657 & 0.733  & 6.64    \\ \hline
$J=5$                   & 0.336 & 0.425 & 0.504 & 0.657 & 0.754 & 9.88    \\ \hline
$J=7$                   & 0.317 & 0.399 & 0.414 & 0.616 & 0.71 & 12.84   \\ \bottomrule
  \end{tabular}
	}
 \label{table:grid search results mimic}
\end{table}

\begin{table}
  \caption{Grid search results on Synthea. $J=\infty$ means skipping the retrieval step.}
  \centering
	\resizebox{0.95\columnwidth}{!}{
  \begin{tabular}{c||c|c|c||c}
    \toprule
Retrieved Documents & Acc@1  & Acc@3  & Acc@5  & \textbf{Avg \#T} \\ \toprule
$J=1$       & 0.562 & 0.6    & 0.6      & 1.125   \\ \hline
$J=2$       & 0.505 & 0.581  & 0.743   & 2.625   \\ \hline
$J=\infty$  & 0.438 & -      & 0.924   & 8    \\ \bottomrule
  \end{tabular}
	}
 \label{table:grid search results synthea}
\end{table}

\begin{table}
  \caption{Results on MIMIC for ablation study with standard deviations across runs. Baseline is the full ReMatch method, names-only refers to skipping the document creation step, and $J=\infty$ means skipping the retrieval step.}
  \centering
	\resizebox{0.95\columnwidth}{!}{
  \begin{tabular}{c||c|c||c}
    \toprule
{Variation}      & Acc@1               & Acc@5  & \textbf{Avg \#T}\\ \toprule
$J=1$ \textit{baseline}           & 0.424 $\pm$ 0.017 & - & 2.44\\ \hline
$J=2$ \textit{baseline}           &    -       & 0.729 $\pm$ 0.023 & 4.68\\ \hline
$J=1$ \textit{names-only} & 0.396 $\pm$ 0.005     & -  & 1.48\\ \hline
$J=2$ \textit{names-only} & -              & 0.503 $\pm$ 0.008 & 3.28\\ \hline
$J=\infty$        & 0.311 $\pm$ 0.0157 & 0.518 $\pm$ 0.03 & 38\\ \bottomrule
  \end{tabular}
	}
 \label{table:ablation study}
\end{table}


\begin{table}
  \caption{Results for MIMIC with guidance provided. Relative improvement over baseline (in percentage) in bold.}
  \centering
	\resizebox{0.95\columnwidth}{!}{
  \begin{tabular}{c||c|c||c}
    \toprule
{Variation}      & Acc@1               & Acc@5 & \textbf{Avg \#T}\\ \toprule
$J=1$            & 0.539/\textbf{20.38\%}  & 0.783/\textbf{10.41\%} & 2.84\\ \hline
$J=2$            & 0.459/\textbf{7.89\%} & 0.765/\textbf{4.86\%} & 4.92\\ \bottomrule
  \end{tabular}
	}
 \label{table: guidance}
\end{table}

\subsection{Results}
\label{subsec:results}
\textbf{ReMatch Results} The results of the grid search for MIMIC can be found in~\cref{table:grid search results mimic}. We found $\{(J=1,\:K=1),\:(J=2,\:K=5)\}$ to be the best balance points between the three optimization dimensions for the MIMIC dataset. Interestingly, accuracy@$K$ seems to perform best for $J\leq2$. This may imply that the retrieval is efficient, and additional tables only add `noise' to the matching prompt.

The results for the grid search for Synthea are presented in\cref{table:grid search results synthea}. While $\{(J=1,\:K=1),\:(J=2,\:K=5)\}$ still appear to work well, for Synthea we found that not using retrieval yielded optimal results for accuracy@5. This may be explained by the small size of the dataset, with only 8 tables in the target. 

The results of the ablation study are shown in~\cref{table:ablation study}. We found that skipping the retrieval step yields inferior results, in alignment with the results from the grid search. We also found that using only the names of the tables and attributes, with no descriptions, results in a significant decrease. 
Additionally,~\cref{table:ablation study} demonstrates that the method's outputs are stable.

Adding guidance to ReMatch indeed improves performance, as can be seen in~\cref{table: guidance}. The largest improvements are for $J=1$, achieving the highest overall scores for MIMIC. This is consistent with the rest of the results, with more efficient retrieval being correlated with better matching accuracy.

\textbf{SMAT Results} \quad
SMAT's performance was validated using the F1-score, with an F1-score of 0.48276 for the 80-20 train-test split on the Synthea dataset, indicating successful model training when compared to the original values reported by Zhang et al. 

\begin{figure}[t!]
\centering
  \begin{subfigure}[t]{0.49\linewidth}
  \centering
\begin{tikzpicture}
  \begin{axis}[
        ybar, axis on top,
        height=4.6cm, width=\linewidth,
        bar width=0.5cm,
        ymajorgrids, tick align=inside,
        major grid style={draw=gray},
        enlarge y limits={value=0.1,upper},
        ymin=0, ymax=0.8,
        axis x line*=bottom,
        axis y line*=left,
        y axis line style={opacity=0},
        tickwidth=0pt,
        enlarge x limits=0.4,
        legend style={
            at={(0.65,1.11)},
            anchor=north,
            legend columns=-1,
            /tikz/every even column/.append style={column sep=0.5cm, font=\tiny}
        },
        symbolic x coords={ACC@1,ACC@5},
       xtick=data,
       xticklabel style={font=\scriptsize, align=center},
       nodes near coords={
        \pgfmathprintnumber[precision=4]{\pgfplotspointmeta}
       },
       every node near coord/.append style={font=\scriptsize}
    ]
   \addplot [fill=blue!40] coordinates {
      (ACC@1,0.077)
      (ACC@5, 0.361) };
    \addplot [fill=green!40] coordinates {
      (ACC@1,0.424)
      (ACC@5, 0.729) };
    \legend{SMAT, ReMatch}
  \end{axis}
  \end{tikzpicture}
  \caption{Performance on MIMIC.}
  \end{subfigure}%
  \hfill
\begin{subfigure}[t]{0.49\linewidth}
\centering
\begin{tikzpicture}
  \begin{axis}[
        ybar, axis on top,
        height=4.6cm, width=\linewidth,
        bar width=0.5cm,
        ymajorgrids, tick align=inside,
        major grid style={draw=gray},
        enlarge y limits={value=0.1,upper},
        ymin=0, ymax=1.0,
        axis x line*=bottom,
        axis y line*=left,
        y axis line style={opacity=0},
        tickwidth=0pt,
        enlarge x limits=0.4,
        legend style={
            at={(0.65,1.11)},
            anchor=north,
            legend columns=-1,
            /tikz/every even column/.append style={column sep=0.5cm, font=\tiny}
        },
        symbolic x coords={ACC@1,ACC@5},
       xtick=data,
       xticklabel style={font=\scriptsize, align=center},
       nodes near coords={
        \pgfmathprintnumber[precision=4]{\pgfplotspointmeta}
       },
       every node near coord/.append style={font=\scriptsize}
    ]
   \addplot [fill=blue!40] coordinates {
      (ACC@1,0.271)
      (ACC@5, 0.694) };
    \addplot [fill=green!40] coordinates {
      (ACC@1,0.562)
      (ACC@5, 0.924) };
    \legend{SMAT, ReMatch}
  \end{axis}
  \end{tikzpicture}
  \caption{Performance on Synthea.}
  \end{subfigure}
  \caption{Comparison of the different models' performance. SMAT was trained and evaluated on 20\%, 80\% of the data, after removing all null mappings. ReMatch was evaluated on the entire dataset, with no guidance, and with nulls. Optimal setup from grid search is shown for ReMatch.}
    \label{fig:comparison of models}
\end{figure}
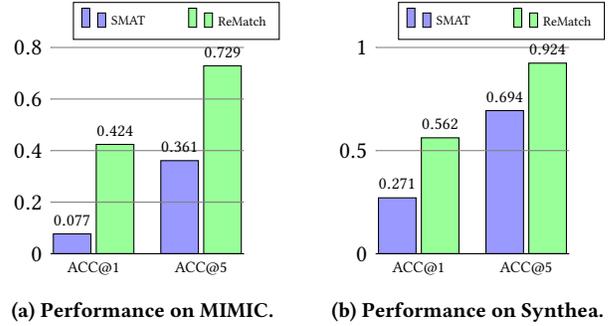

When using 80\% of the data for training, SMAT achieved a perfect accuracy@5 score (accuracy@5 = 100\%) on the remaining 20\% of the attributes. Unfortunately, in real-world scenarios only a small portion of mappings can be available, since mapping 80\% manually removes the need for automatic mapping. In the much more realistic split, with only 20\% of the data, SMAT achieved significantly lower results than ReMatch on both datasets, as is emphasized in~\cref{fig:comparison of models}. Unlike ReMatch, the results were significantly higher on Synthea, when compared to MIMIC dataset, for both splits. This finding is aligned with our expectations about the difficulty of MIMIC, and demonstrates the robustness of ReMatch. 

 \begin{table}
  \caption{Results for SMAT on both datasets.}
  \centering
	\resizebox{\columnwidth}{!}{
  \subfloat[MIMIC]{
  \begin{tabular}{c||c|c|c}
    \toprule
{Variation}      & Acc@1               & Acc@5  & F1 \\ \toprule \bottomrule
SMAT 80-20       & 0.454 & 1.0    & 0.102\\ \hline
SMAT 20-80       & 0.077 & 0.361 & 0.084 \\  \bottomrule
  \end{tabular}}
\quad
\subfloat[Synthea]{
  \begin{tabular}{c||c|c|c}
    \toprule
{Variation}      & Acc@1               & Acc@5  & F1 \\ \toprule \bottomrule
SMAT 80-20       & 0.875  & 1.0    & 0.483\\ \hline
SMAT 20-80       & 0.271 & 0.694 & 0.241 \\ \bottomrule
  \end{tabular} }
	}
 \label{table:ditto and smat MIMIC}
\end{table}

\section{Conclusions}
\label{sec:conclusions}
In this work we introduced ReMatch, a scalable and effective method for matching schemas using retrieval-enhanced LLMs, designed to complement and aid human matchers throughout their work. ReMatch avoids the need for predefined mapping, any model training, or access to data in the source database. Instead, it exploits the generative abilities of LLMs to perform semantic ranking between two schemas, in alignment with a human matching process. We also provided a new large dataset that will hopefully aid further research.

In future work, we plan to include the explicit use of type constraints, foreign keys, and primary keys, as well as enhanced guidance mechanisms and enrichment of table and column descriptions using the source schema data, where accessible without privacy or security constraints. 


\bibliographystyle{ACM-Reference-Format}
\bibliography{refs}

\end{document}